\documentclass{iau}
\usepackage{graphicx}

\title[Milky Way's nuclear cluster and BH] 
{The Milky Way's nuclear star cluster and massive black hole}

\author[Rainer Sch{\"o}del]   
{Rainer Sch{\"o}del$^1$
}

\affiliation{$^1$Instituto de Astrof{\'i}sica de Andaluc{\'i}a (CSIC), \\ Glorieta de la Astronom{\'i}a s/n,
ES-18008, Granada, Spain \\ email: {\tt rainer@iaa.es}
}

\pubyear{2015}
\volume{312}  
\pagerange{--}
\jname{Star clusters and black holes in galaxies acrosss cosmic time}
\editors{R. Spurzem, F. Liu, S. Li \& Y. Meiron, eds.}
\begin{document}

\maketitle

\begin{abstract}
Because of its nearness to Earth, the centre of the Milky Way is the only galaxy nucleus in which we can study the characteristics, distribution, kinematics, and dynamics of the stars on milli-parsec scales. We have accurate and precise measurements of the Galactic centre's central black hole, Sagittarius A*, and can study its interaction with the surrounding nuclear star cluster in detail. This contribution aims at providing a concise overview of our current knowledge about the Milky Way's central black hole and nuclear star cluster, at highlighting the observational challenges and limitations, and at discussing some of the current key areas of investigation. 
\end{abstract}

\firstsection 
\section{Introduction}

The paradigm that massive black holes (MBHs) are located in the centres of all major galaxies has been firmly established over the past decades (e.g., \cite[G{\"u}ltekin \etal\, 2009]{Gultekin09}). Moreover, the recent detection of an MBH in an ultra-compact dwarf galaxy by \cite[Seth \etal\ (2014)]{Seth14}  indicates that these objects may be present in almost all galaxy types. In addition to MBHs, the majority of galaxies of all types contain stellar nuclei in the form of nuclear star clusters (NSCs). NSCs have similar sizes as globular clusters, but contain several times their mass. They thus belong to the densest stellar systems in the Universe (for an overview, see, e.g., \cite[B{\"o}ker 2010]{Boeker10}, or \cite[Sch{\"o}del \etal\ 2014b]{Schoedel14b}). The study of galactic nuclei is of great interest for astrophysics in general because it touches a wide range of topics, such as General Relativity (GR), dense N-body dynamics, star formation in  extreme environments or the accretion history of MBHs. Unfortunately, the large distances of extragalactic nuclei mean that we can resolve only relatively large physical scales in them (at best a few 0.1 pc in the nearest systems). This means that the observed light per resolution element arises from tens of thousands to millions of stars and the radius of influence, where the MBH dominates stellar dynamics, is barely resolved. 

The Milky Way appears to be a relatively normal  barred spiral galaxy. Since its nucleus is located at a distance of only about 8 kpc from Earth, it provides us with a unique opportunity to study galactic stellar nuclei and MBHs. In the case of the Galactic centre (GC), we can resolve linear scales on the order of milli-parsecs (mpc) in the near-infrared (NIR) and thus examine individual stars and their kinematics and dynamics. Hence, the GC is of fundamental importance for investigating questions such as the validity of GR near an MBH, the interaction of stars with an MBH, the initial mass function in galaxy nuclei, or the existence of stellar cusps around MBHs through accurate and precise quantitative measurements. For further in-depth reading about most of the topics discussed here, I recommend the recent detailed review article about the GC by \cite[Genzel \etal\ (2010)]{Genzel10}, as well as the shorter review by \cite[Sch{\"o}del \etal\ (2014b)]{Schoedel14b}, that is mainly focused on the Milky Way's NSC. 


\section{Observational constraints}

While the Milky Way offers a unique template for the study of galactic nuclei on the one hand, there exist, on the other hand, significant constraints for observational studies of the GC. Our line-of-sight through the Galactic disc implies that interstellar extinction toward the Milky Way's nucleus is extreme. With $A_{V}\gtrsim30$\,mag, studies at visual wavelengths are all but impossible. Even in the NIR, at wavelengths around $2.2\,\mu$m (the so-called $K$-band), extinction still amounts to 2-5\,mag (see \cite[Nishiyama \etal\ 2008]{Nishiyama08} or \cite[Fritz \etal\ 2011]{Fritz11}). The presence of molecular clouds in the central few hundred parsecs of the Milky Way induces the additional difficulty that interstellar extinction varies significantly on angular scales of only a few arcseconds (\cite[Sch{\"o}del \etal\ 2010]{Schoedel10}).  This makes even rough stellar classification through broad-band photometry, e.g., distinguishing between cool giants and massive main sequence stars, very challenging.

The sheer number of stars results in high surface number densities of at least a few (at $\sim$100\,pc from the center) up to several tens (in the central parsec) of stars per square arc-second. As a consequence, crowding limits the completeness of star counts to relatively bright magnitudes ($K\lesssim15$) and can induce significant astrometric and photometric bias in  seeing-limited (resolution $\sim0.5"$ at $2.2\,\mu$m) imaging from the ground. For the central parsec of the GC, even the resolution offered by the Hubble Space Telescope (HST) is insufficient to overcome crowding. To overcome crowding, the resolution offered by an 8-10m-class telescope supported by adaptive optics (resolution $\sim0.06"$) is needed. In any case, the 50\% completeness limit for source detection lies at magnitudes as bright as $K\approx18-19$ in the central parsec of the GC. The detection of solar mass main sequence stars in this region ($K\approx21$, taking into account distance and extinction) will require the angular resolution of telescopes of the 30m-class. 

\section{Sagittarius\,A*: The Milky Way's central black hole}

Stellar proper motion and line-of-sight velocity measurements carried out since the mid-1990s (see \cite[Eckart \& Genzel 1996]{Eckart96} or \cite[Ghez \etal\ 1998]{Ghez98}) have resulted in the accurate measurement of a large number of individual orbits of stars around the radio source Sagittarius\,A* (\cite[Gillessen \etal\ 2009]{Gillessen09} and \cite[Ghez \etal\ 2008]{Ghez08}). The currently existing data can be fitted accurately with Keplerian orbits and require a mass of $\sim$$4\times10^{6}\,M_{\odot}$ to be concentrated within a radius of  $\lesssim0.6\,$mpc of Sagittarius\,A* (Sgr\,A*). The resulting high mass density in combination with radio-to-X-ray measurements of the (very low) emission from this location make the black hole hypothesis the only one that can currently satisfyingly explain all observational data. Although the term Sgr\,A* refers, strictly speaking, to the electromagnetic radiation released by hot plasma close to the Milky Way's central MBH, it is frequently used as a name for the putative black hole itself. 

Stellar orbits have also allowed us to measure the distance of Sgr\,A*, which is $\sim$8\,kpc. The combined statistical and systematic uncertainties of the mass and distance of the black hole are currently already $<10\%$ and $<5\%$, respectively, and will further improve with continued monitoring of stellar orbits in the future.

Stars with orbital periods less than 20\,yr are termed short-period stars. They are of special importance because their orbital parameters can be determined with high accuracy within reasonable time. They are therefore of paramount importance to probe the gravitational potential around Sgr\,A* and thus to determine the amount of extended mass around the black hole, that may be present in the form of stellar remnants, and, at the same time, to test the validity of General Relativity. Several short-period stars are needed to break model degeneracies (see \cite[Meyer \etal\ 2012]{Meyer12}). The star with the currently shortest-known period orbits Sgr\,A* in just $11.5$\,years (\cite[Meyer \etal\ 2012]{Meyer12}). The star with the most accurately and precisely measured orbit is S2/S0-2. It will pass through peri-centre again in 2018 and will then provide us with an opportunity to detect the transverse-Doppler effect and gravitational redshift terms of special and general relativity (for an overview of this topic, see, e.g., \cite[Genzel \etal\ 2010]{Genzel10} or \cite[Sch{\"o}del \etal\ 2014a]{Schoedel14a})

\section{Nuclear star cluster: Morphology and kinematics}

One of the most complete existing works on the stellar structures at the GC is the study by \cite[Launhardt \etal\ (2002)]{Launhardt02}. They show that the central hundreds of parsecs are dominated by the so-called {\it nuclear bulge} (NB), that is composed of a stellar disc (scale height $\lesssim45\,$pc, radius $\sim$230\,pc) and a compact NSC. The total stellar mass of the NB is about $1.4\times10^{9}\,M_{\odot}$. Strong UV-radiation and the stellar luminosity function indicate significant recent star formation, in particular toward the NSC. As concerns the properties of the NSC, a limitation in this study and other, previous and later studies is, however, that they are significantly affected by low angular resolution and/or the strong and variable interstellar extinction. The shape of the NSC was assumed as spherical without being able to test this assumption. Two recent works have made substantial progress in this respect, but with different methodologies. \cite[Fritz \etal\ (2014)]{Fritz14} have corrected stellar number counts for extinction and corrected areas with extreme extinction by assuming symmetry with respect to the Galactic plane and axis. \cite[Sch{\"o}del \etal\ (2014a)]{Schoedel14a} used IRAC/Spitzer mid-infrared images around $3-5\,\mu$m because interstellar extinction reaches a minimum in this region (\cite[Fritz \etal\ 2011]{Fritz11}). They thus did not have to assume any intrinsic symmetry but could instead directly demonstrate the point-symmetry of the NSC with respect to Sgr\,A*. Both studies agree in their findings, which is encouraging, given the completely different methodologies applied. The NSC is found to be intrinsically flattened. It is aligned with the Galactic plane, although a small misalignment ($\lesssim10\,^{\circ})$ cannot be excluded.

Both mentioned studies generally agree in the measured parameters of the Milky Way's NSC. According to \cite[Sch{\"o}del \etal\ (2014a)]{Schoedel14a}, the nuclear cluster of the Milky Way
\begin{enumerate}
\item is precisely centred on Sgr\,A* (uncertainty $<0.2$\,pc on large scales; actually, higher angular resolution measurements of a smaller area show that the uncertainty is only on the order of $0.02$\,pc, see \cite[Sch{\"o}del \etal\ 2007]{Schoedel07});
\item has a ratio of minor to major axis of $0.71\pm0.02$;
\item can be described adequately by a S{\'e}rsic law with an index $n=2.0\pm0.2$;
\item has a half light radius of $4.2\pm0.4$\,pc; and
\item has a luminosity of $4.1\pm0.4\times10^{7}\,L_{\odot}$ and a mass of $2.5\pm0.4\times10^{7}\,M_{\odot}$.
\end{enumerate}

The most recent studies of the overall kinematics of the Milky Way's NSC were carried out by \cite[Feldmeier \etal\ (2014)]{Feldmeier14} and \cite[Chatzopoulos \etal\ (2015)]{Chatzopoulos15} (see also \cite[Fritz \etal\ 2014]{Fritz14}). Also these studies use completely different methodologies. While \cite[Chatzopoulos \etal\ (2015)]{Chatzopoulos15} use the proper motions and line-of-sight velocities of thousands of individual stars, obtained from observations with high angular resolution, \cite[Feldmeier \etal\ (2014)]{Feldmeier14} analyse a seeing-limited spectroscopic slit drift-scan map  of the NSC. Encouragingly, the basic results are again in agreement. 
\begin{enumerate}
\item  The kinematics of the NSC are consistent with its flattening. 
\item The NSC rotates in parallel to Galactic rotation. Its rotation velocity approaches asymptotically a values of $30-40$\,km\,s$^{-1}$ at distances $r\gtrsim4\,$pc.
\item The NSC can be described approximately by an isotropic rotator model (\cite[Chatzopoulos \etal\ 2015]{Chatzopoulos15}). 
\item The kinematically derived mass is consistent with the photometrically derived mass by \cite[Sch{\"o}del \etal\ (2014a)]{Schoedel14a}.
\item The stellar mass-to-light ratio is $0.76\pm0.18\,M_{\odot}/L_{\odot}$ at $\sim$$2.2\,\mu$m  (\cite[Chatzopoulos \etal\ 2015]{Chatzopoulos15}) and $0.56\pm0.26\,M_{\odot}/L_{\odot}$ at $\sim$$4.5\,\mu$m (\cite[Feldmeier \etal\ 2014]{Feldmeier14}).
\item The kinematic analysis suggests a misalignment of the NSC major axis by $\sim$9$^{\circ}$ with respect to the Galactic plane. 
\item There are indication of a separate kinematic component at distances $r=1-2\,$pc from Sgr\,A*, whose position angle appears to be offset by $\sim$$90^{\circ}$ from the overall cluster rotation.  
\end{enumerate}

The kinematic substructure and the overall misalignment of the NSC kinematic axis with the Galactic plane may indicate the residuals of distinct accretion events. Investigating this evidence further may provide hints to the formation history of the NSC. 

Finally, it is important to note that kinematic modelling of the NSC almost always consistently underestimates the mass of the central MBH by factors of $30-50\%$. The studies by \cite[Chatzopoulos \etal\ (2015)]{Chatzopoulos15} and, to a lesser degree, by \cite[Sch{\"o}del \etal\ (2009)]{Schoedel09} and \cite[Do \etal\ (2013)]{Do13} are the only ones that derive BH masses largely consistent with the estimates from stellar orbits. However, it must be pointed out that these publications were produced when the correct solution was already known. \cite[Fritz \etal\ (2014)]{Fritz14} list and discuss and analyse quantitatively possible sources of bias, among them the effects of anisotropy and the flat stellar core of the cluster. They conclude that no effect by itself appears to be sufficient to explain the low derived BH mass in the Jeans modeling. A combination of various effects and/or the necessity for improved modeling may provide the answer. Data sets such as the pseudo integral-field slit-scan data of \cite[Feldmeier \etal\ (2014)]{Feldmeier14} are of great value to study the possible systematic effects that lead to erroneous BH masses. This is of particular importance with respect to spectroscopic studies of extragalactic MBHs, which may suffer similar biases. 

\section{Is there a stellar cusp around Sgr\,A*?}

The formation of a stellar cusp in a relaxed stellar cluster around a MBH is a robust result of theoretical stellar dynamics, which predicts a stellar density increase in the form of a power-law within the sphere of influence of the black hole. The latter is generally defined by the radius within which the stellar mass corresponds to twice the black hole mass. Depending on the properties of the stellar cluster, in particular the number ratio between heavy and light stars, the stellar density within the sphere of influence will settle to a density distribution that can be described by $\rho\propto r^{-\gamma}$, with different values of $\gamma$ for stars of different mass. \cite[Alexander \& Hopman (2009)]{Alexander09} show that  in the so-called weak mass segregation regime $\gamma\approx7/4$ is valid for the heavy stars and $3/2<\gamma<7/4$ for the light stars. In the strong segregation regime, which is considered probable for the Milky Way's NSC, the density distribution of the rare massive stellar objects can be described by $2\lesssim\gamma\lesssim11/4$  and the one of the lighter stars by $3/2\lesssim\gamma\lesssim7/4$ (see also \cite[Preto \& Amaro-Seoane 2010]{Preto10}).

Searching for stellar cusps around extragalactic MBHs is a very difficult task because the related small angular scales of their radius of influence mean that one can only study the light density profile. The latter can, however, easily be biased by the presence of a small number of bright giants/supergiants as well as by interstellar extinction and recent star formation events near the MBH. Conclusive tests require number density counts or very careful removal of the influence of bright stars (see discussion in \cite[Sch{\"o}del \etal\ 2007]{Schoedel07}). With current technology, the GC is therefore the only reliable target to test the predictions of theoretical stellar dynamics on cusp formation. 

Within the radius of influence of Sgr\,A*, about 2-4\,pc (\cite[Feldmeier \etal\ 2014]{Feldmeier14}, \cite[Chatzopoulos \etal\ 2015]{Chatzopoulos15}), the projected stellar number density follows a density law of about $\rho\propto R^{-0.8}$, where $R$ is the projected radius. This is consistent with a three-dimensional stellar cusp of $\rho\propto r^{-1.8}$, i.e., close to the predicted values. However, the number density of the stars old enough to be dynamically relaxed has been found to be almost flat within a projected radius of roughly 0.5\,pc around Sgr\,A* (\cite[Buchholz \etal\ 2009]{Buchholz09}; \cite[Do \etal\ 2009]{Do09}; \cite[Bartko \etal\  2010]{Bartko10}). This means that the NSC appears to be characterised by a core instead of a cusp around Sgr\,A*. This surprising finding is inconsistent with the theoretical predictions. 

There exist a range of models that try to explain the observed absence of a stellar cusp in the immediate vicinity of Sgr\,A*. The first class of explanations assumes that the density distribution of the observed stars is representative for the entire population of the NSC. In that case, the relaxation time in the NSC may simply be too long for the cusp to have formed (\cite[Merritt 2010]{Merritt10}) or the cusp may have been destroyed (e.g., \cite[Merritt \& Szell 2006]{Merritt06}). The second class assumes that the cusp is invisible, i.e., that the observable giant stars are no adequate tracers of the underlying stellar distribution. This could happen if collisions destroy the envelopes of the giants and render them invisible (e.g., \cite[Dale \etal\ 2009]{Dale09}; \cite[Amaro-Seoane \& Chen 2014]{Amaro14}). Another explanation is provided  by \cite[L{\"o}ckmann \etal\ (2010)]{Lockmann10}: Continuous star formation over the Galaxy's lifetime results in the formation of stellar black holes within the NSC that migrate towards the centre due to dynamical friction and push out lighter stars to greater distances, thus turning the cusp of visible stars into a core. 

At the moment it is not clear whether the absence of evidence for the stellar cusp is evidence for its absence. The observational constraints (see above) imply that we can currently only detect a small fraction of the stellar population of the NSC, that is, giants, supergiants, and massive, short-lived main and post-main sequence stars. The detection of main-sequence stars of (sub-)solar mass will require the sensitivity and, in particular,  angular resolution of a 30m-class telescope. Also, future high-precision observations of stellar orbits may reveal the presence of an extended dark mass component around Sgr\,A* (see, e.g., \cite[Weinberg \etal\ 2005]{Weinberg05}; \cite[Perets \etal\ 2009]{Perets09}).

\section{Star formation near Sgr\,A*}

Several studies over the past decades have tried to infer the star formation history of the Milky Way's NSC, a difficult task given the challenging observational limitations. The most recent study was carried out by \cite[Pfuhl \etal\ (2011)]{Pfuhl11}. It was mainly based on spectroscopy of 450 cool giant stars within a projected distance of 1\,pc of Sgr\,A*, but also included some information on intermediate-mass main sequence stars from very sensitive spectroscopic observations. \cite[Pfuhl \etal\ (2011)]{Pfuhl11} find that about $80\%$ of the stellar mass had already formed about 5\,Gyr ago. About $1-2$\,Gyr ago there was an apparent minimum in the star formation rate, which increased again during the last few hundred Myrs. The data appear to indicate that the bulk of the NSC's stars formed with a canonical Chabrier/Kroupa initial mass function (see also \cite[L{\"o}ckmann \etal\ 2010]{Lockmann10}).

Much observational and theoretical effort has been focused on understanding the properties of the most recent star formation event in the GC. It is traced by roughly 180 O/B super giants and main sequence stars as well as Wolf-Rayet stars (e.g., \cite[Levin \& Beloborodov  2003], \cite[Paumard \etal\ 2006]{Paumard06}; \cite[Bartko \etal\  2009]{Bartko09}, \cite[Bartko \etal\  2010]{Bartko10}, \cite[Lu \etal\ 2009]{Lu09}, \cite[Lu \etal\ 2013]{Lu13}). Almost all of these stars are located within 0.5\,pc in projection from Sgr\,A* and a significant fraction of them rotate in a clockwise disk around the MBH. Their density increases toward the MBH and the disk-like pattern of their dynamics is consistent with their formation in a formerly existing dense gas disk around Sgr\,A*. The viability of this scenario has been confirmed by theoretical models (see, e.g., \cite[Bonnell \& Rice 2008]{Bonnell08}, and discussions in \cite[Levin \& Beloborodov  2003], \cite[Lu \etal\ 2009]{Lu09}, \cite[Bartko \etal\  2010]{Bartko10}, or \cite[Genzel \etal\ 2010]{Genzel10}). The age of this star formation event (if it was a single one), is estimated to between $2.5$ to $5.8$\,Myr (\cite[Lu \etal\ 2013]{Lu13}). Intriguingly, this recent star formation recent event was almost certainly characterised by a top-heavy IMF (see, e.g.,  \cite[Nayakshin \& Sunyaev 2005]{Nayakshin05}, \cite[Bartko \etal\  (2010)]{Bartko10}, \cite[Lu \etal\ 2013]{Lu13}), in contrast to the finding that most stars in the NSC formed with a standard IMF.

\section{Summary}

Because of extreme interstellar extinction and high stellar surface density the centre of the Milky Way poses, on the one hand,  unique observational challenges. On the other hand, it is the only nucleus of a quiescent spiral galaxy nucleus that we can resolve observationally on scales of milli-parsecs and plays therefore a key role as a template, where to test many of our theoretical ideas. The existence of a central  black hole at the GC has been established with high confidence through the measurements of stellar orbits and radio-to-X-ray observations of its electromagnetic counterpart, Sagittarius\,A*. Uncertainties on its mass and distance are already $\lesssim10\%$ and can be expected to reach percent-level in the next decade. This will open the door to using the GC as a calibrator for cosmic distance measurements. 

With a half-light radius of about 4\,pc, a mass of $\sim$$2.5\times10^{7}\,M_{\odot}$, and its complex stellar population, the Milky Way's NSC appears to be very similar to extragalactic nuclear clusters. It is truly central, both in morphology and kinematics, rotates in parallel to overall Galactic rotation, and is significantly flattened along the Galactic plane. Apparent kinematic substructures and the possible kinematic misalignment of the NSC with the Galactic plane may be remnants of individual accretion events, that may have contributed to building the cluster. A point of potential importance for measuring the masses of extragalactic MBHs is that the mass of Sgr\,A* has almost always been consistently under-estimated when applying Jeans modeling to kinematic data.

One of the key questions about galactic nuclei that we can test in the Milky Way's NSC is the presence of a stellar cusp around the MBH. Contrary to robust theoretical predictions, the NSC shows a core-like structure in the central parsec. It is not clear whether the cusp is indeed absent, i.e.\ has not yet had the time to form or was destroyed, or is just invisible because it is composed of dark remnants or because the tracer stars (giants) have been rendered invisible through collisions. Future measurements of stellar dynamics and sensitive, high angular resolution observations with 30m-class telescopes may help us to better understand the mystery of the missing cusp.

 Although we can still not reconstruct the Milky Way's NSC's  formation history, it appears to be clear that the majority of its stars formed many Gyrs ago with a standard IMF. Recent star formation occurred close to Sgr\,A* a few Myr ago. The stars probably formed in a dense gas disc around the black hole with a top-heavy IMF. 

 \acknowledgments 
The research leading to these results has received funding from the European Research Council under the European Union's Seventh Framework Programme (FP/2007-2013) / ERC Grant Agreement n.\ 614922, and by grants AYA2010-17631 and AYA2012-38491-CO2-02, cofunded with FEDER funds, of the Spanish Ministry of Economy and Competitiveness.

\end{document}